%% file: manuscript.tex
\documentclass[a4paper, 12pt, review]{elsarticle} 

\usepackage[latin1]{inputenc}
\usepackage[T1]{fontenc}
\usepackage{ae}
\usepackage[automark]{scrpage2}
\usepackage{array}
\usepackage[intlimits]{amsmath}
\usepackage{amsfonts, amssymb, amsthm, amstext}
\usepackage{graphicx}
\usepackage{hyperref}
\usepackage[usenames, dvipsnames]{color}

\newcommand{\prespectrometer}{pre-spec\-tro\-meter}

\begin{document}

\begin{frontmatter}
  \title{Background due to stored electrons following nuclear decays in the KATRIN spectrometers and its impact on the neutrino mass sensitivity}

  \author[address1]{S. Mertens\corref{cor}}
  \ead{susanne.mertens@kit.edu}
  
  \author[address1]{G. Drexlin} 
  \author[address1,address2]{F.M. Fr{\"a}nkle }
  \author[address4]{D. Furse}
  \author[address1,address3]{F. Gl{\"u}ck}
  \author[address1]{S. G{\"o}rhardt}
  \author[address1]{M. H{\"o}tzel}
  \author[address1]{W. K{\"a}fer}
  \author[address1]{B. Leiber}
  \author[address1]{T. Th{\"u}mmler}
  \author[address1]{N. Wandkowsky}
  \author[address1]{J. Wolf}

  \address[address1]{KCETA, Karlsruhe Institute of Technology, 76131 Karlsruhe, Germany}
  \address[address2]{Department of Physics, University of North Carolina, Chapel Hill, NC, USA}
  \address[address3]{Research Institute for Nuclear and Particle Physics, Theory Dep., Budapest, Hungary}
  \address[address4]{Massachusetts Institute of Technology, Cambridge, MA, USA}
  \cortext[cor]{Corresponding author}

\include{abstract}
\end{frontmatter}

\section{Introduction} 
\label{Introduction} 
The Karlsruhe Tritium Neutrino (KATRIN) experiment is a next generation, large-scale, tritium $\beta$-decay experiment currently under construction at the Karlsruhe Institute of Technology (KIT); it will prospectively start taking data in 2015. KATRIN is designed to measure the effective electron anti-neutrino mass $\text{m}_{\overline{\nu}_e}$, defined as  
\begin{equation}
  \text{m}_{\overline{\nu}_e} = \sqrt{\sum\limits_{i=1}^{3}|\text{U}_{\text{e}i}|^2\cdot \text{m}_i^2}, 
\end{equation}
where $\text{U}_{\text{e}i}$ denotes the Pontecorvo-Maki-Nakagawa-Sakata leptonic mixing matrix and $\text{m}_i$ are the neutrino mass eigenstates~\cite{OttenWeinheimer}. The design sensitivity of KATRIN is 200~meV at 90\% confidence level~\cite{DesignReport}.

The experiment will use a model-independent technique based on the kinematics of tritium $\beta$-decay. It will analyze the shape of the electron energy spectrum in a narrow region close to the tritium decay endpoint at $E_0 = 18.6$~keV. A non-zero neutrino mass reduces the maximum energy of the electron and changes the shape of the tritium $\beta$-spectrum in the immediate vicinity of the endpoint. To reach the neutrino mass sensitivity, several criteria including high energy resolution, high signal count rates and low background must be fulfilled.

\begin{figure}[]
\begin{center}
\includegraphics[width = \textwidth]{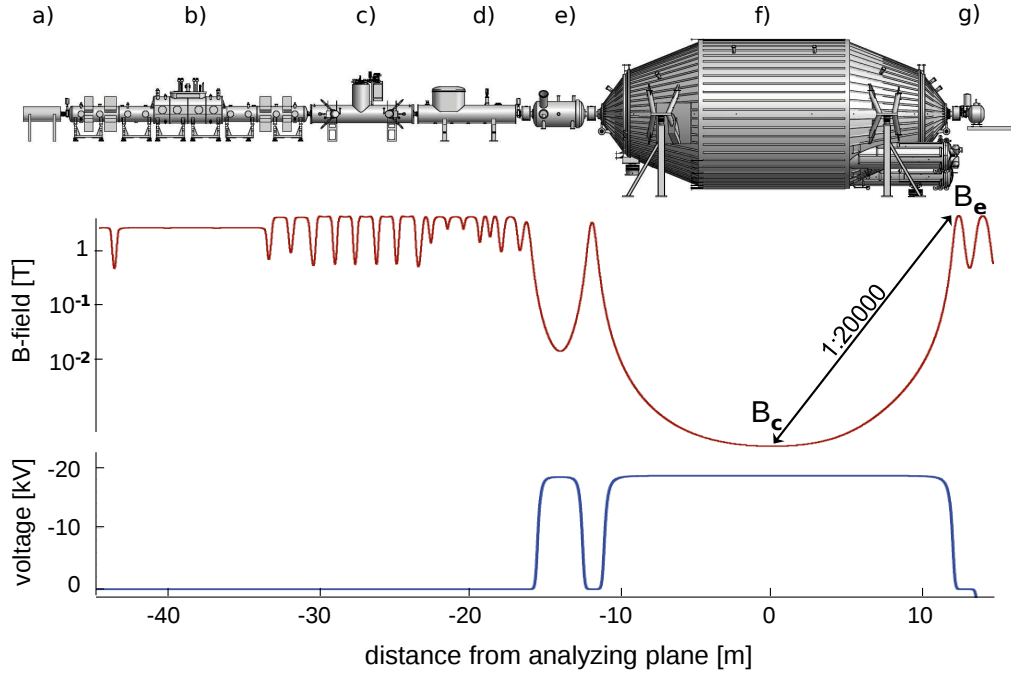}
\caption{\textbf{KATRIN experimental setup with main components} a: rear section, b: windowless gaseous tritium source, c: differential pumping section, d: cryogenic pumping section, e: \prespectrometer{}, f: main spectrometer, g: focal plane detector. Below, the magnetic field and the electric potential along the beam axis are displayed. In both spectrometers the MAC-E filter principle is applied: As the electric potential is increased to $\text{U}_{\text{ret}}=-18.6$~kV to filter the $\beta$-electrons, the magnetic field drops from $\text{B}_{\text{e}}=6$~T to $\text{B}_{\text{c}}=3\cdot10^{-4}$~T, which collimates the electrons into a parallel beam with a flux of $\Phi=191~\text{T}\text{cm}^2$.}
\label{fig:KATRIN}
\end{center}
\end{figure}

In the 70~m long KATRIN setup (shown in figure \ref{fig:KATRIN}) a windowless gaseous tritium source (WGTS) of high stability and luminosity is combined with a large electrostatic retarding spectrometer of unsurpassed resolution~\cite{DesignReport}. A magnetic guidance system adiabatically transports the electrons created in the tritium source towards the spectrometer where the energy analysis takes place. The spectrometer, working as an electrostatic filter, transmits only those electrons which have sufficient energy to overcome the retarding potential. The transmitted electrons are then counted at a detector. By measuring the count rate for different filter voltages, the shape of the integrated energy spectrum can be determined.

Since the spectrometer section must be essentially tritium-free, the tritium flow is reduced from the WGTS injection rate of $1.8$~$\text{mbar}\cdot\ell/\text{s}$ down to a value of $10^{-14}$~$\text{mbar}\cdot\ell/\text{s}$ at the end of the transport section. This unprecedentedly large suppression factor will be achieved by a combination of differential (DPS) and cryogenic pumping (CPS) elements~\cite{DPS2FMeas,DPS2FSim,CPSMeas,CPSSim}.

From the electron creation in the WGTS until the energy analysis in the central analyzing plane of the main spectrometer, the magnetic field drops by four orders of magnitude, collimating the electron momenta via the magnetic gradient force. This combination of Magnetic Adiabatic Collimation with Electrostatic filter, called the MAC-E filter principle, further described in section~\ref{MACE}, allows for large solid angle acceptance, combined with high energy resolution \cite{MACE1, MACE2}. 

In this paper we perform a detailed investigation of a novel background source arising from stored multi-keV electrons produced in $\beta$-decays of tritium and secondary processes occurring during $\alpha$-decays of the radon isotopes ${}^{219,220}$Rn in the volume of the large main spectrometer. Due to the known magnetic bottle characteristics of a MAC-E filter for light charged particles, the electrons arising from nuclear decays inside the spectrometer volume are magnetically stored. With storage times of up to several hours, these particles can significantly enhance the background level via ionization of residual gas. 

Despite the huge tritium retention factor, careful radio assays and use of low-activity components, the $\nu$-mass measurements can be seriously disrupted by even single tritium $\beta$-decays or $\alpha$-decays of short-lived Rn-isotopes. This is due to exceedingly long storage times and the large number of background events resulting from one decay. We know of no other case in astroparticle physics experiments where a single nuclear decay can continuously influence the measurements over a time period of several hours.

Here we outline a detailed model of these processes validated by measurements at the much smaller  \prespectrometer{}~\cite{Fraenkle,Nancy} and use it to predict background rates and characteristics of the final KATRIN setup. Based on these results, we investigate implications on the neutrino mass sensitivity and demonstrate that the original KATRIN setup described in~\cite{DesignReport} would result in background levels exceeding the required limits. To mitigate these problems we finally propose important design refinements and novel active background reduction methods.

This paper is structured as follows: Section~\ref{MACE} briefly describes the fundamental principles of the MAC-E filter as they pertain to the background described in this work. Section~\ref{Tools} gives a brief introduction to the software used to perform the MC simulations. In section~\ref{Mechnism}, the mechanisms of background production will be outlined. In section~\ref{MS}, we discuss the expected event rates based on this background model for different vacuum scenarios in the spectrometer section, while the impact of this new background source on the KATRIN sensitivity will be discussed in section~\ref{Sensitivity}. 

\section{MAC-E filter principle of KATRIN}
\label{MACE}
The WGTS produces electrons at a rate of~$\sim 10^{11}$~Hz which are emitted isotropically and guided along magnetic field lines towards the spectrometer section, itself consisting of a smaller \prespectrometer{} providing the option to filter out low-energy electrons, and a larger main spectrometer for precision filtering. The magnetic guidance of the electrons through the spectrometer is provided by a system of three superconducting coils (see figure~\ref{fig:KATRIN}) and an external air coil system surrounding the main spectrometer. The retarding potential in both spectrometers is provided by inner electrodes constructed from wires, which allow for a $10^{-6}$ precision of the filter potential~\cite{Valerius}. The kinetic energy $E_\text{kin}$ of electrons entering the spectrometer section may be decomposed as 
\begin{equation}
	E_\text{kin} = E_{\perp} + E_{\parallel},
\end{equation}
where $E_{\perp}$ denotes the energy associated with the cyclotron motion and $E_{\parallel}$ corresponds to longitudinal motion along a magnetic field line. Of $E_\text{kin}$ only $E_{\parallel}$ is analyzed by the electrostatic filter. To achieve both high count rates and superior energy resolution, the initial $E_{\perp}$ component must be transformed into $E_{\parallel}$ on the way to the central analyzing plane. This is achieved by the MAC-E filter principle, where, in the case of KATRIN, the magnetic field drops by four orders of magnitude from the entrance (or exit) of the spectrometer to its center (see figure~\ref{fig:KATRIN}). By extending the reduction of the magnetic field strength over a length of about 10~m, a very smooth change of the magnetic field is assured, resulting in a fully adiabatic motion of the electrons. Due to this adiabaticity, the orbital magnetic moment $\mu$ of electrons is conserved. To first order, $\mu$ is given by
\begin{equation}
	\mu \approx \frac{E_{\perp}}{|\vec{B}|} \approx \text{const.}
\end{equation}    
The reduction of the magnetic field strength thus transforms the transversal energy at the edge ($E_{\perp}^\text{e}$) almost completely into parallel energy at the center of the spectrometer ($E_{\parallel}^\text{c}$)
\begin{equation}
	E_{\parallel}^\text{c} = E_{\text{kin}}^\text{c} - E_{\perp}^\text{c} = E_{\text{kin}}^\text{c} - E_{\perp}^\text{e}\frac{|\vec{B}|^\text{min}}{|\vec{B}|^\text{max}} = E_{\text{kin}}^\text{c} - E_{\perp}^\text{e}\cdot5\cdot10^{-5},
\end{equation}
where quantities considered at the edge of the spectrometer carry superscript e, and those considered at center carry superscript c. It is $E_{\parallel}^\text{c}$ which is analyzed by the electrostatic filter.

The MAC-E filter technique is as yet the most sensitive technique used in direct neutrino mass experiments~\cite{OttenWeinheimer}, and a central design feature of the KATRIN experiment. The drawback, however, is that this magnetic field configuration inherently forms a magnetic bottle for light charged particles, since both ends of the spectrometer work as magnetic mirrors~\cite{MagneticMirrorBook,MagneticMirror1,MagneticMirror2}.

\section{Simulation tools}
\label{Tools}

The main principles of the MAC-E filter and its application in the KATRIN experiment can be understood analytically via the adiabatic approximation. However, in order to illuminate the role of the MAC-E filter as it applies to stored particle backgrounds and the complex, non-adiabatic situations these entail, a precise and fast computational tool is required. The tasks of such a tool include the calculation of electromagnetic fields and particle trajectories to machine precision. This tool is provided by the simulation software \textsc{Kassiopeia}~\cite{Kassiopeia, Kassiopeia2}, which has been developed over the past years by the KATRIN collaboration.

The trajectory calculations of \textsc{Kassiopeia} are based on explicit Runge-Kutta methods described in~\cite{RungeKutta, RungeKutta2, RungeKutta3}. Electric and magnetic field calculations are performed via the zonal harmonic expansion~\cite{FerencEl, FerencMag}. In the case of electric fields, computations are carried out using the boundary element method~\cite{BEM}. Elastic, electronic excitation and ionization collisions of electrons with molecular hydrogen are included in the simulations; they are based on data and calculations in~\cite{scattering1, scattering2, scattering3, scattering4, scattering5, scattering6, scattering7, scattering8, scattering9}. The field, tracking and scattering simulations originate from FORTRAN and C codes developed between 2000 and 2008 by one of us (F.{}~G.).  

In the framework of the investigations presented below, \textsc{Kassiopeia} was equipped with a selection of event generators including the $\beta$-decay of tritium and $\alpha$-decays of different radon isotopes. The simulation of tritium $\beta$-decay is implemented using Fermi's theory of weak interactions~\cite{Fermi34, Sheldon}. Here we make use of a detailed tritium generator which includes the final state distribution of tritium~\cite{FinalStates1, FinalStates2} and radiative corrections~\cite{RadiativeCorrections} to the $\beta$-spectrum, while shake-off effects at low energies are not yet included~\cite{ShakeOffT2}.

The modeling of electrons produced in radon $\alpha$-decays includes processes described in more detail below, such as the creation of shake-off electrons produced in the initial $\alpha$-decay and conversion, shell reorganization and Auger electrons produced in the decay of the daughter polonium isotopes. The simulation of these processes is based on data in the code Penelope~\cite{Penelope} and the literature~\cite{ConversionDataRn220, ConversionDataRn219, ShakeOff, KShakeOffRadon, ShellReorganization, LMShakeOffRadon}.

The software has been validated by a number of associated measurements mostly performed as test experiments for KATRIN~\cite{Prall} and cross-checked with other methods~\cite{PartOpt, Simion} as well as analytic calculations~\cite{Jackson,ValiMagZonal1, ValiMagZonal4, ValiMagZonal5, ValiEl1, BEM2}. 
The radon event generator in particular is validated through comparisons to \prespectrometer{}~\cite{Fraenkle, Nancy} and independent measurements~\cite{Rn220ChargeDist}.

\section{Background production mechanism}
\label{Mechnism}
In this section, the basic ingredients of the background production mechanism will be outlined. 
First, the nuclear decays of tritium and radon are examined in detail as sources of primary high-energy electrons in the keV range. Second, the electrons' dynamical behavior in a MAC-E filter and the mechanism of particle trapping and the conditions under which such storage may occur will be discussed. The final part of this section relates these processes to the observed background rates.

\subsection{Nuclear decays as source of high-energy electrons} 
\label{ElectronSources}
The main source of keV-range primary electrons are nuclear decays. Of particular concern for the KATRIN experiment are tritium $\beta$-decays and $\alpha$-decays of the short-lived radon isotopes ${}^{219,220}$Rn. 

As a central design requirement of KATRIN~\cite{DesignReport}, only an exceedingly small fraction of the order of $10^{-14}$ of the tritium molecules injected into the WGTS will reach the spectrometer section. A small number of these molecules will decay there before being pumped out, thereby generating electrons with a continuous spectrum of up to about $E_{0}$.

Electrons in a similar or higher energy range can be produced following nuclear $\alpha$-decays (the primary $\alpha$-particle as well as fluorescence X-rays are of no concern here). Due to the large pumping speed of the turbomolecular pumps (TMPs) connected to the main spectrometer volume and the correspondingly short pumping times (about 360~s), only short-lived radon isotopes are of importance here. Therefore ${}^{222}$Rn emanation, relevant for underground experiments like Gerda~\cite{Gerda}, is not an issue for KATRIN due to its long lifetime ($\tau_{222} = 5.51$~d).
On the other hand, the number of ${}^{219}$Rn ($\tau_{219} = 5.71$~s) and ${}^{220}$Rn ($\tau_{220} = 80.2$~s) decays in the sensitive volume is not reduced significantly by pumping. Thus, these $\alpha$-decays generate a background source which is distributed homogeneously over the entire spectrometer volume ($\text{V}_{\text{MS}} = 1290~\text{m}^{3}$).

${}^{219}$Rn arises from the ${}^{235}$U actinide decay chain and emanates in small quantities primarily from the non-evaporable getter (NEG) material~\cite{Fraenkle} used for pumping the spectrometers; for details of the Rn emanation from NEG material see~\cite{Fraenkle, Susanne, Florian}. The stainless steel inner surfaces of the main and \prespectrometer{} as well as auxiliary equipment such as ceramic insulators, glass windows, vacuum gauges and thermocouples emanate ${}^{220}$Rn arising from the ${}^{232}$Th decay chain. 
The electron energy spectrum resulting from ${}^{219}$Rn and ${}^{220}$Rn $\alpha$-decays can be attributed to the processes of internal conversion, shake-off, shell reorganization and the Auger effect.

\begin{figure}
  \centering
  \includegraphics[height = 7cm]{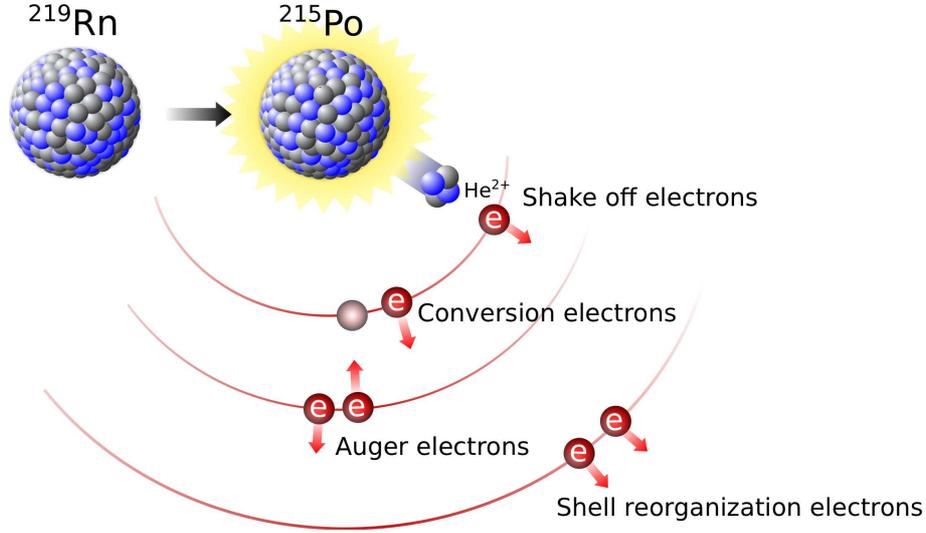}
  \caption{\textbf{Visualization of the different electron production mechanisms in ${}^{219}$Rn $\alpha$-decay}. Details of the mechanisms are explained in the text.}
  \label{fig:RadonDecayElectrons}
\end{figure}

Figure~\ref{fig:RadonDecayElectrons} is a sketch of the electron generation processes following radon $\alpha$-decay. ${}^{219}$Rn (${}^{220}$Rn) decays into excited ${}^{215}\text{Po}^{*}$ (${}^{216}\text{Po}^{*}$) states which then decay within a few picoseconds. If the wave function of a shell electron is non-vanishing at the nucleus, a conversion electron can be emitted in the de-excitation process, instead of radiating gammas \cite{ConversionDataRn219, ConversionDataRn220}. This process is dominant for heavy nuclei, and in the case of polonium de-excitation, conversion electrons can reach energies of up to $E_{\text{e}} = 450$~keV. 

The emitted $\alpha$-particle can directly knock out shake-off electrons from the atomic shells. These electrons reach energies of up to $E_{\text{e}} = 80$~keV~\cite{ShakeOff, LMShakeOffRadon, KShakeOffRadon}. Additionally, the emission of the $\alpha$-particle results in a sudden, non-adiabatic change of the nuclear potential, which leads to the emission of predominantly two low-energy shell reorganization electrons from the outer shells which share an energy of about $E_{\text{e}} = 250$~eV~\cite{ShellReorganization}.

Subsequent to shake-off and conversion electron processes, which both may leave vacancies in the electron shell, Auger electrons can be emitted. The latter process often involves cascades of relaxations \cite{Penelope}, releasing multiple electrons with energies of up to $E_{\text{e}} = 20$~keV.

\begin{figure}
\centering
\includegraphics[width=\textwidth]{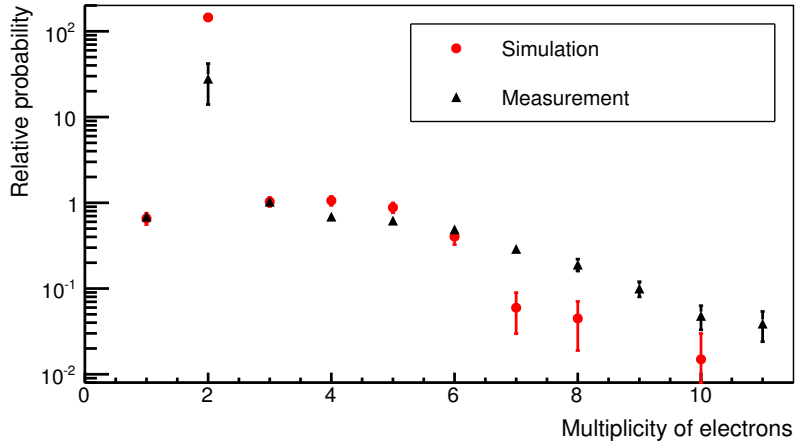}
\caption{\textbf{Multiplicity of electrons following ${}^{220}$Rn $\alpha$-decays.} The MC simulation with \textsc{Kassiopeia} based on the model described here is compared to the independent measurement in~\cite{Rn220ChargeDist}, demonstrating the soundness of our background model.}
\label{fig:Multiplicity}
\end{figure}

As an example of the complexity of the processes involved, figure~\ref{fig:Multiplicity} shows the electron multiplicity of ${}^{220}$Rn $\alpha$-decay as simulated for this investigation and previously measured in an independent work~\cite{Rn220ChargeDist}. The simulation and measurements agree well, demonstrating the basic validity of our event generators. These generators are described in more detail in~\cite{Nancy}.

\subsection{The main spectrometer as magnetic mirror trap}
\label{MirrorTrap}
Due to the operating principle of the MAC-E filter, an electron produced in the center of the spectrometer is accelerated towards regions of low electric potential at the ends of the spectrometer, thereby moving from a region of low to high magnetic field. Consequently, its longitudinal energy $E_{\parallel}$ is transformed into transversal energy $E_{\perp}$. Depending on the starting angle and energy of the electron, the kinetic energy can be completely transformed into transversal energy so that the electron is magnetically trapped. 

Nevertheless, there are situations under which the storage conditions are broken: 
\begin{itemize}
	\item Below a certain minimum transversal starting energy $E^{\text{min}}_{\perp}$, the electron cannot be magnetically trapped, since the acceleration by the electric field is too strong. In case of the main spectrometer, this minimal energy is $E^{\text{min}}_{\perp} = 0.93$~eV.
	\item Above a certain transversal starting energy $E^{\text{max}}_{\perp}$, the electron's cyclotron radius becomes larger than the radius of the main spectrometer, and therefore the electron directly hits the wall. For the reference field $\text{B}_{\text{min}} = 3\cdot10^{-4}$~T and the dimensions of the main spectrometer ($\text{\o} = 10$~m), this corresponds to $E^{\text{max}}_{\perp} = 180$~keV.
	\item  The motion of high-energy electrons in low magnetic fields can be non-adiabatic. Accordingly, the transformations of $E_{\perp}$ into $E_{\parallel}$ and vice versa are no longer proportional to the change of the magnetic field, i.e.{} the angle of the momentum vector to the magnetic field line changes randomly. Therefore, non-adiabatic motion allows the electron to eventually escape the magnetic mirror trap.
\end{itemize}

\begin{figure}[htbp]
\begin{center}
\includegraphics[width = \textwidth]{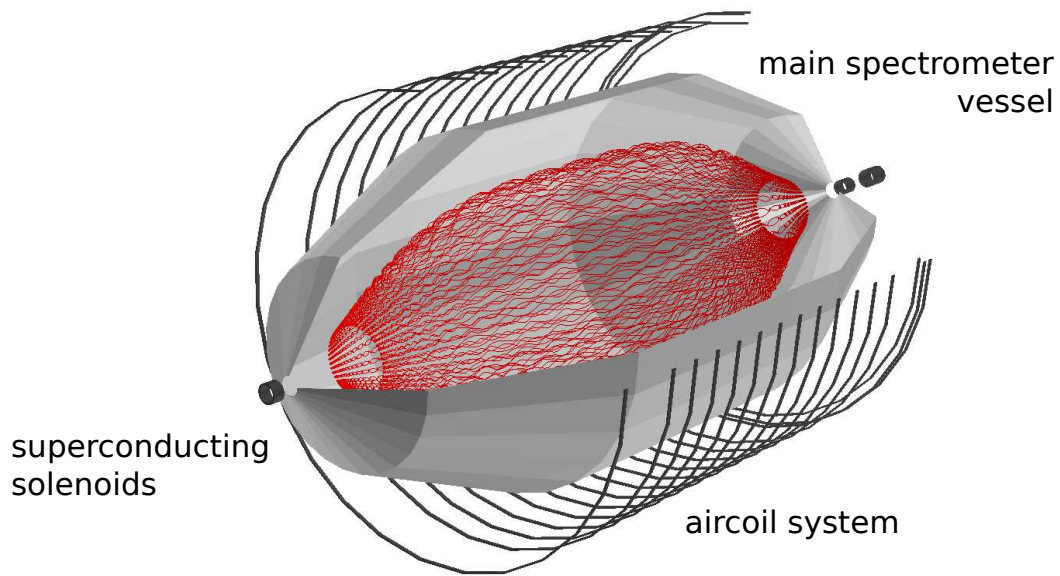}
\caption{\textbf{Calculation of the trajectory of a stored 3~keV electron in the main spectrometer with KASSIOPEIA.} An electron produced via a nuclear decay can be trapped due to the magnetic mirror effect. Its motion is a superposition of cyclotron motion, axial motion and azimuthal magnetron drift. Here, the trajectory is shown for a time period about $100~\mu$s, whereas the actual storage time is much longer.}
\label{fig:trapped_radon}
\end{center}
\end{figure}

\subsection{Background production of stored high-energy electrons} 
\label{Cooling}
A primary high-energy (keV-range) electron which is trapped in the magnetic bottle slowly cools down via ionization and electronic excitation collisions with residual gas molecules. Other energy loss mechanisms such as elastic scattering and emission of synchrotron radiation only play a minor role. Due to the excellent ultra high vacuum (UHV) conditions ($\text{p} = 10^{-11}$~mbar), collisions are rare, allowing a single electron to be stored for several hours. The hundreds of secondary electrons generated by ionizing collisions are mainly low-energy and typically leave the spectrometer on a rather short time scale of minutes. Accelerated by the retarding potential they hit the detector and thus produce a background in the narrow energy interval of the signal $\beta$-decay electrons (the energy region-of-interest ROI is from $15-21$~keV~\cite{DesignReport}).

The total number of secondary electrons $N_{\text{s}}$ for a fixed primary energy given approximately by 
\begin{equation}
  N_{\text{s}}(E_{\text{prim}}) \approx \frac{E_{\text{prim}}}{\omega},
\end{equation}
where $\omega = 37$~eV  denotes the average energy of ion electron pair creation off $\text{H}_2$ for electrons in the keV-range~\cite{Eloss} and $E_{\text{prim}}$ represents the primary starting energy.
For a realistic calculation of $N_{\text{s}}$ in our specific case, however, the following corrections need to be taken into account: 
\begin{itemize}
\item The high-energy secondary electrons themselves can be stored again and produce tertiary electrons,
\item at very high energies, electrons may leave the magnetic trap before being fully cooled down due to non-adiabatic effects,
\item stored electrons additionally lose energy by emitting synchrotron radiation.
\end{itemize}

\begin{figure}
\centering
\includegraphics[width=\textwidth]{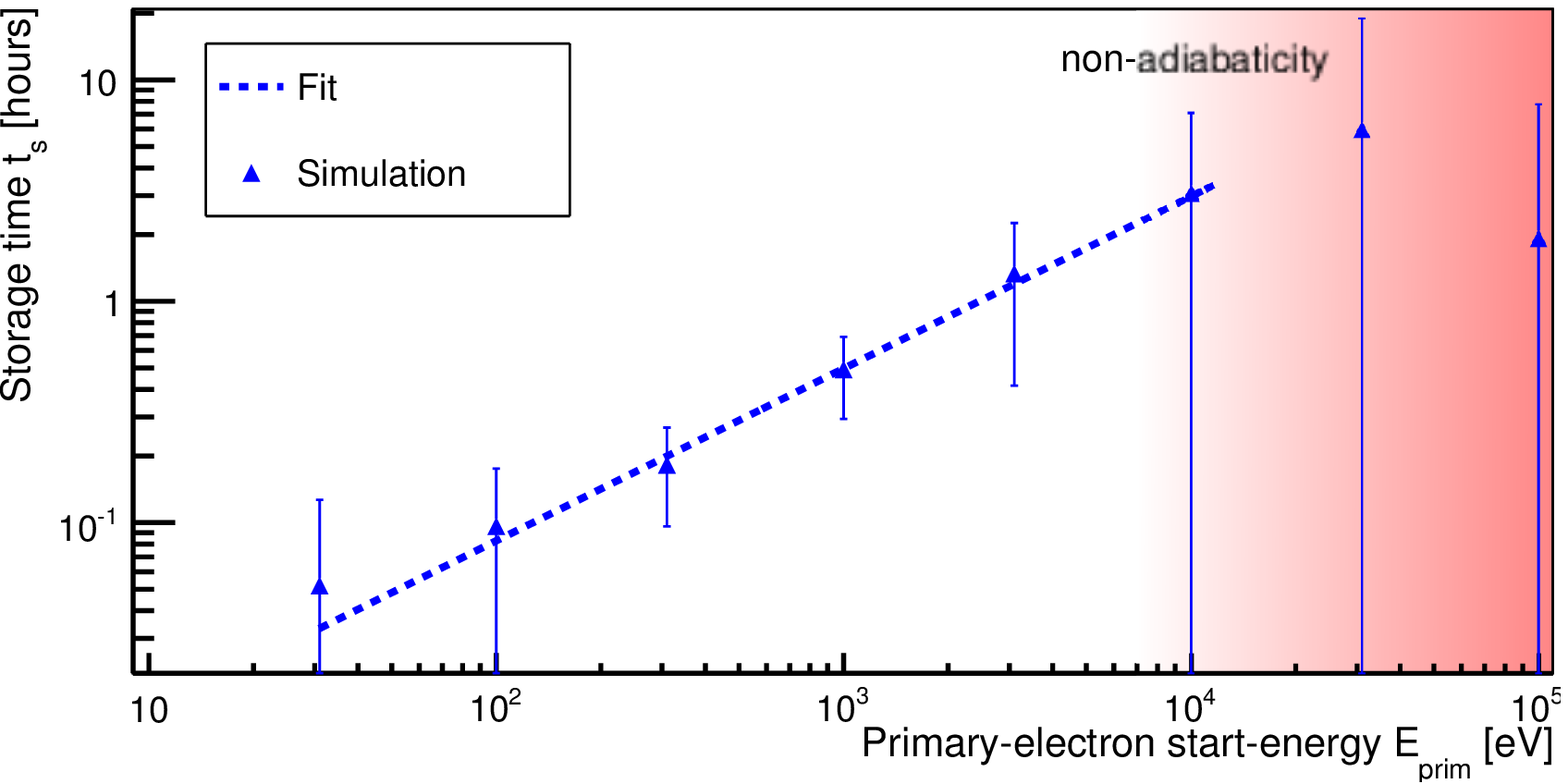}
\includegraphics[width=\textwidth]{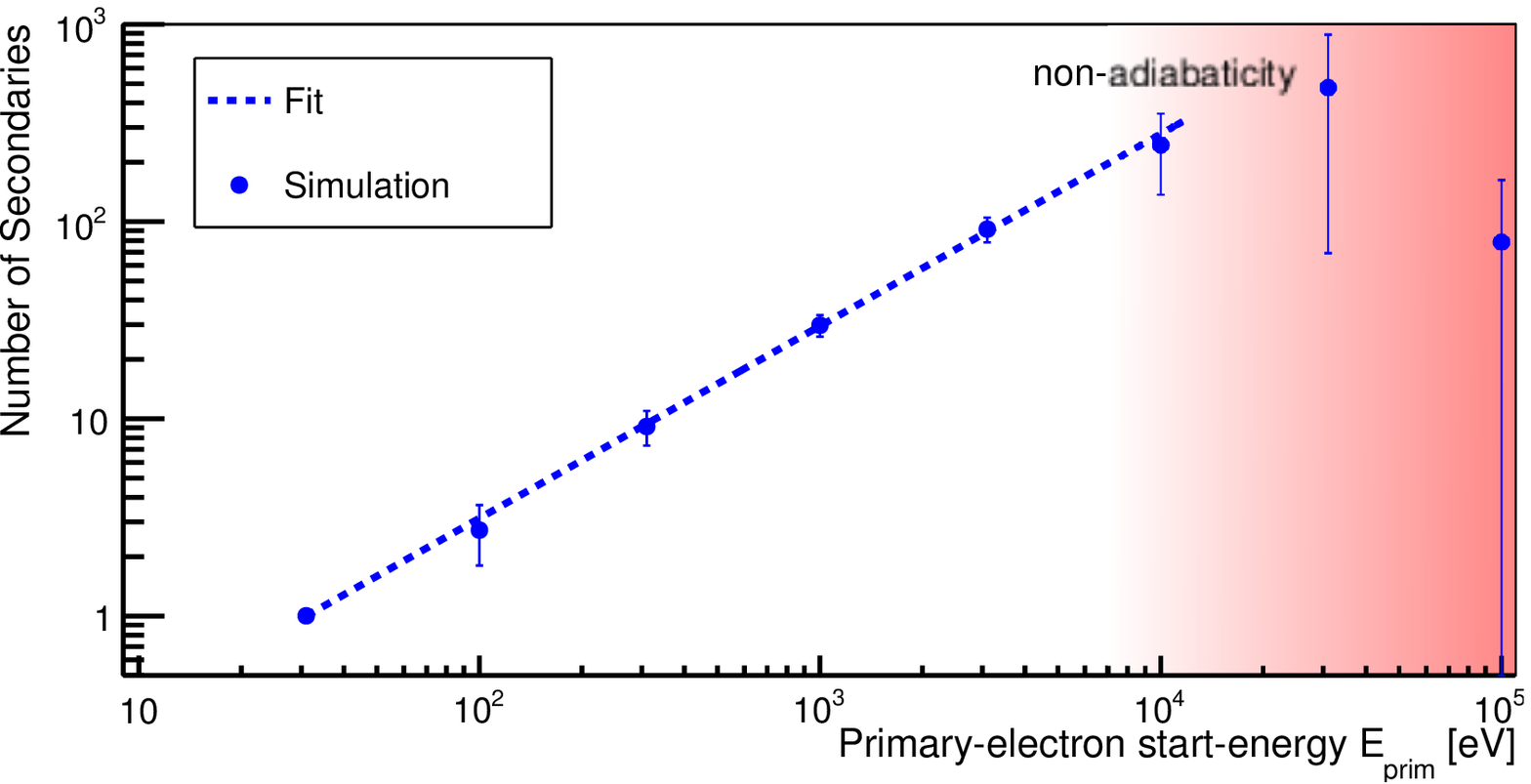}
\caption{\textbf{Average storage time $t_{\text{s}}$ (top) and average number of secondary electrons $N_{\text{s}}$ (bottom) as a function of the primary electron starting energy $E_{\text{prim}}$.} The error bars signify the standard deviation away from this average. The \textsc{Kassiopeia} simulations reveal a power law behavior scaling with $t_{\text{s}}\sim E_{\text{prim}}^{(0.78\pm0.25)}$ and $N_{\text{s}}\sim E_{\text{prim}}^{(0.97\pm0.02)}$ for $E_{\text{prim}} < 30$~keV. Above 30~keV the effects of non-adiabatic motion become dominant, as indicated by the shaded area, reducing the average storage time of the primary electron and the number of secondaries, respectively.}
\label{fig:duration_nsec_energy}
\end{figure}

To incorporate these effects, we have carried out extensive simulations with the \textsc{Kassiopeia} simulation package, investigating the important parameters of primary storage time $t_{\text{s}}$ (defined as the time between the creation of the primary and the end of its trajectory) and number of secondary electrons $N_{\text{s}}$ as a function of $E_{\text{prim}}$. The results are shown in figure~\ref{fig:duration_nsec_energy}. In these simulations, eight different energies were selected in a range between 10~eV~--~100~keV. For each of these values, $10^3$ electrons were started isotropically in the main spectrometer. The simulation takes into account elastic scattering, excitation and ionization on $\text{H}_2$ at a pressure of $\text{p} = 10^{-11}$~mbar, as well as non-adiabatic effects and synchrotron radiation. The average computation time of a single 10~keV stored electron and all its secondaries is $t_{\text{comp}} \approx 8\cdot10^4$~s on an Intel Xeon X5550 2.67~GHz processor. 

The trajectory of each electron (and all secondary electrons) was computed until it 
\begin{itemize}
	\item leaves the spectrometer through the entrance or exit port,
	\item or hits the spectrometer electrodes or vessel wall,
	\item or was cooled down below the ionization threshold $E_{\text{thres}} = 15$~eV.
\end{itemize}
This cut-off parameter is motivated by the minor influence of electron ionization interactions below $E_{\text{thres}}$ on our results. A detailed investigation of the processes below $E_{\text{thres}}$ will be described in a separate publication~\cite{LowEnergy}.

The results of our \textsc{Kassiopeia} simulations reveal a clear correlation of both background parameters with $E_{\text{prim}}$: a higher $E_{\text{prim}}$ implies a longer storage time (up to 10~h) and a higher multiplicity of secondary electrons. As a generic example, a 10~keV electron leads to the creation of~$\sim 300$ secondaries in a time period of 3~h, corresponding to a background rate of $\text{r}_{\text{B}} = 30$~mHz in the energy ROI. For energies  $E_{\text{prim}}<30$~keV, the average number of secondaries is a good means of estimation of the primary electron starting energy $E_{\text{prim}}$. 

At energies above $E_{\text{prim}}\sim30$~keV, the effects of non-adiabatic motion become more prominent. The large storage times and the considerable number of secondary electrons underline the importance of the detailed investigations below.

\section{Expected background rates at the main spectrometer}
\label{MS}
In this section, we use the background model described above to estimate the actual background rate to be expected at the final KATRIN setup. In doing so we examine four different vacuum strategies for the KATRIN spectrometer section to minimize the background rates.

\subsection{Sources of radon and tritium}
\label{RadonTritiumSources}
The \prespectrometer{} measurements~\cite{Fraenkle} initially revealed the dominant background source to be ${}^{219}$Rn emanation from the 90~m of NEG strips (1.8~kg NEG material). Further sources of ${}^{219,220}$Rn emanation were identified to be specific vacuum gauges and sensor instrumentation. After removal of the getter pump and the auxiliary components, a small number of events with radon-like characteristics were still observed, which we attribute to radon emanation from the inner surface of the stainless steel walls. 

The number of radon decays expected in the main spectrometer can be extrapolated from these measurements.  
In the main spectrometer a much larger NEG pump with 3000~m NEG strips (60~kg NEG material) will be used. The ${}^{219}$Rn emanation for this batch was reduced by a factor of two through a special production process. However, since the decay series is not in secular equilibrium, the radon production rate increases slowly over time at a rate of $0.3~\text{Bq}/(\text{kg}\cdot\text{yr})$~\cite{Joachim}. 

As mentioned before, the remaining radon events after removal of the getter and auxiliary components are assumed to be caused by emanation from the walls. For the following discussion, we assume that this radon emanation rate scales with the respective spectrometer vessel surface $\text{A}_i$ (surface area of the \prespectrometer{}: $\text{A}_{\text{PS}} \approx 25~\text{m}^2$, main spectrometer: $\text{A}_{\text{MS}} \approx 690~\text{m}^2$). We assume a negligible emanation of radon isotopes from vacuum gauges and sensor instrumentation, as well as from the structural materials of the inner electrode system~\cite{Valerius}.  

To calculate the number of tritium decays in the main spectrometer, we use the maximum allowed tritium flow from the WGTS into the \prespectrometer{}, approximately  $Q_{\text{T}_2}^{\text{CPS $\rightarrow$ PS}} \approx 10^{-14}~\text{mbar}\cdot\ell/\text{s} = 2.5\cdot 10^{5}$~molecules/s, as detailed in~\cite{DesignReport}. Additionally, due to the large number of adsorption/desorption processes in the transport section, we note that the gas flow will be a mixture of hydrated tritium (HT) and other hydrogen isotopologues, in the context of this work, however, we only consider $\text{T}_2$.

\subsection{The vacuum system of the KATRIN spectrometer section}
\label{VacuumSystem}	
The vacuum system of the KATRIN spectrometer section~\cite{Vacuum} is based on two pumping strategies: TMPs to pump out noble gas atoms, such as radon, and secondly NEG pumps for pumping out hydrogen isotopologues, including tritium. Accordingly, for the NEG system, there is an inherent trade-off between increased tritium pumping capacity and enhanced radon emanation.

Specifically, the vacuum system of the spectrometer section consists of six TMPs~\cite{TMP} installed in pairs at the three pump ports at the detector-facing end of the main spectrometer and two smaller TMPs at the \prespectrometer{} pump ports. The pump ports in the main spectrometer (\prespectrometer{}) are additionally equipped with 3000~m (180~m) of NEG strips~\cite{NEG, NEGSticking}. 

To reduce the number of radon atoms reaching the main spectrometer volume from the NEG strips, $\text{LN}_2$-cooled cryo-baffle systems were installed in front of each of the three pump ports. The excellent performance of this method to shield the sensitive flux tube of a spectrometer from the pump port which houses the NEG strips has been demonstrated previously~\cite{Stefan}.

The presence of the baffle, however, also results in a decrease of the hydrogen (tritium) pumping speed. As will be shown in the following, an optimum solution is found for a configuration with 250~m additional getter strips in the \prespectrometer{} and the installation of cryo-baffles in all three pump ports of the main spectrometer.

Table~\ref{tab:Vacuum} summarizes the sources and corresponding reduction rates of radon atoms and tritium molecules for specific layouts of the KATRIN vacuum system.

\begin{table}
\caption{This table displays all important input parameters for the calculation of radon and tritium decay rates. We list the emanation rates of radon in the main spectrometer from 3000~m getter strips with and without the baffle installed ($E^{\text{MS}}_{\text{3~km NEG + baf.}}$, $E^{\text{MS}}_{\text{3~km NEG}}$) as well as from the wall ($E^{\text{MS}}_{\text{Wall}}$). As another source, the inflow of radon and tritium from the \prespectrometer{} for the cases of 180~m getter ($\text{Q}^{\text{PS}\rightarrow\text{MS}}_{\text{180~m NEG}}$) and 180~m plus additional 250~m getter strips ($\text{Q}^{\text{PS}\rightarrow\text{MS}}_{\text{add. 250~m NEG}}$) are given. Furthermore, the effective pumping speeds of the TMPs ($S^{\text{MS}}_{\text{6 TMPs}}$), and the NEG pump with baffle ($S^{\text{MS}}_{\text{3~km NEG + baf.}}$) and without ($S^{\text{MS}}_{\text{3~km NEG}}$) are listed for tritium and radon. Finally, we list the radioactive decay constants ($\lambda$). For all calculations a sticking coefficient of 0.8 for the cryo-baffle is assumed. The tritium inflow rates are to be understood as upper limits and are therefore given without an error estimation. The errors on radon emanation rates are propagated from measurements at the \prespectrometer{}.}
\centering
{\footnotesize
\begin{tabular*}{\textwidth}{lllll}
\hline
i (isotope) & $\text{T}_2$ & ${}^{219}\text{Rn}_{\text{NEG}}$ & ${}^{219}\text{Rn}_{\text{Wall}}$ & ${}^{220}\text{Rn}_{\text{Wall}}$\\
\hline\hline 
  \multicolumn{5}{l}{\textbf{Emanation and inflow rates in $\lbrack \frac{1}{s} \rbrack$}} \\
\hline
$E^{\text{MS}}_{\text{3~km NEG}}$(i)  & 0 & $0.12\pm0.03$ & $0.03\pm0.03$ & $0.08\pm0.06$ \\
$E^{\text{MS}}_{\text{3~km NEG + baf.}}$(i)  & 0 & 0 & $0.03\pm0.03$ & $0.08\pm0.06$ \\
$E^{\text{MS}}_{\text{Wall}}$(i) & 0 & 0 & $(4\pm4) \cdot 10^{-5}$  & $(12\pm8) \cdot 10^{-5}$ \\
$\text{Q}^{\text{PS}\rightarrow\text{MS}}_{\text{180~m NEG}}$(i) & 3110 & $(27\pm6.7) \cdot 10^{-5}$ & $(3.7\pm3.7) \cdot 10^{-5}$ & $(40\pm27) \cdot 10^{-5}$ \\
$\text{Q}^{\text{PS}\rightarrow\text{MS}}_{\text{add. 250~m NEG}}$(i) & 1289 & $(64\pm16)  \cdot 10^{-5}$ & $(3.7\pm3.7) \cdot 10^{-5}$ & $(40\pm27) \cdot 10^{-5}$ \\
\hline
\hline
  \multicolumn{5}{l}{\textbf{Effective pumping speeds $S$ in $\lbrack \frac{l}{s} \rbrack$}} \\
\hline
$S^{\text{MS}}_{\text{6 TMPs}}$(i) & 3510 & 3500 & 3500  & 12010\\
$S^{\text{MS}}_{\text{3~km NEG}}$(i) & 577350 & 0 & 0 & 0\\
$S^{\text{MS}}_{\text{3~km NEG + baf.}}$(i) & 259810 & 901860 & 901860 & 899800\\
\hline
\hline
  \multicolumn{5}{l}{\textbf{Radioactive decay constants in $\lbrack \frac{1}{s} \rbrack$}} \\
\hline
$\lambda_{\text{i}}$ & $3.58 \cdot 10^{-9}$ & 0.175 & 0.175 & $1.25 \cdot 10^{-2}$ \\
\hline
\end{tabular*}}
\label{tab:Vacuum}
\end{table}

\subsection{Calculation of the expected decay rates}
\label{DecayRateCalculations}
The time dependent number of radon atoms $N^{\text{MS}}(\text{Rn})$ obeys the following differential equation: 
  \begin{footnotesize}
  \begin{equation} \label{eq:ODE}
  \frac{\text{d}N^{\text{MS}}(\text{Rn})}{\text{d}t} =  -\underbrace{\lambda_{\text{Rn}}\cdot N^{\text{MS}}(\text{Rn})}_{\text{total decay rate}} - \underbrace{\frac{N^{\text{MS}}(\text{Rn})}{V_{\text{MS}}}\cdot S^{\text{MS}}(\text{Rn})}_{\text{pump out rate}} + \underbrace{Q^{\text{PS}\rightarrow \text{MS}}(\text{Rn})}_{\text{inflow from PS}} + \underbrace{E^{\text{MS}}(\text{Rn})}_{\text{emanation rate}},
  \end{equation}
  \end{footnotesize}
where $\lambda_{\text{Rn}}$ denotes the radioactive decay constant of the corresponding radon isotope, $V_{\text{MS}}$ stands for the volume of the main spectrometer and $S^{\text{MS}}(\text{Rn})$ is given by the sum of the available pumping systems for a particular radon isotope in the corresponding UHV scenario.
The back-flow of radon from the main spectrometer to the \prespectrometer{} can be neglected. 
In equilibrium we expect
  \begin{footnotesize}	  
  \begin{equation} \label{eq:equi}
  \frac{\text{d}N^{\text{MS}}(\text{Rn})}{\text{d}t} = 0
  \end{equation}
  \end{footnotesize}
and one finds the number of radon isotopes $N^{\text{MS}}(\text{Rn})$ to be
  \begin{footnotesize}
  \begin{equation}
  N^{\text{MS}}(\text{Rn}) = \left[Q^{\text{PS}\rightarrow \text{MS}}(\text{Rn}) + E^{\text{MS}}(\text{Rn})\right]\cdot \frac{V_{\text{MS}}}{\lambda_{\text{Rn}}\cdot V_{\text{MS}} + S^{\text{MS}}(\text{Rn})}. \label{equ:n_ms}
  \end{equation}	  
  \end{footnotesize} 
The number of tritium molecules $N^{\text{MS}}(\text{T}_2)$ in the main spectrometer is described similarly by 
  \begin{footnotesize}
  \begin{equation}
  \frac{\text{d}N^{\text{MS}}(\text{T}_2)}{\text{d}t}= -\underbrace{\lambda_{\text{T}_2}\cdot N^{\text{MS}}(\text{T}_2)}_{\text{total decay rate}} - \underbrace{\frac{N^{\text{MS}}(\text{T}_2)}{V_{\text{MS}}}\cdot S^{\text{MS}}(\text{T}_2)}_{\text{pump out rate}} + \underbrace{Q^{\text{PS} \rightarrow \text{MS}}(\text{T}_2)}_{\text{inflow from PS}},
  \end{equation}	  
  \end{footnotesize}
where $\lambda_{\text{T}_2}$ denotes the radioactive decay constant of $\text{T}_2$ and $S^{\text{MS}}(\text{T}_2)$ is given by the sum of the available pumping systems for $\text{T}_2$ for a specific UHV scenario.

Using the input parameters summarized in table~\ref{tab:Vacuum}, the decay rates are computed for four different scenarios outlined in table~\ref{tab:Scenarios}:
\begin{itemize}
\item Scenario 1 completely avoids background from NEG correlated radon decay activity,
\item Scenario 2 primarily reduces background arising from tritium decay,
\item Scenario 3 optimally reduces both background rates arising from tritium and radon decays, see figure~\ref{fig:barplot}, 
\item Scenario 4 uses no additional getter in the \prespectrometer{}. This scenario will be realized at the start-up of the spectrometer test measurements in 2012. 
\end{itemize}

\begin{table}
\caption{Description of scenarios. The scenarios are differentiated in their selection of the amount of NEG strips in the \prespectrometer{} and main spectrometer as well as the usage of $\text{LN}_2$ cooled baffle.}
\centering
{\footnotesize
\begin{tabular*}{\textwidth}{p{2.cm}p{2.3cm}p{2.7cm}p{2.3cm}p{2.3cm}}
\hline
  & \multicolumn{2}{l}{\textbf{\prespectrometer{}}} & \multicolumn{2}{l}{\textbf{main spectrometer}} \\
\hline\hline
  & 180~m NEG & add.{} 250~m NEG & 3000~m NEG & $\text{LN}_2$ baffle\\
\hline
Scenario 1 & $-$ & $-$ & $-$ & $-$ \\
Scenario 2 & $\checkmark$ & $\checkmark$ & $\checkmark$ & $-$ \\
Scenario 3 & $\checkmark$ & $\checkmark$ & $\checkmark$ & $\checkmark$ \\
Scenario 4 & $\checkmark$ & $-$ & $\checkmark$ & $\checkmark$ \\
\hline
\end{tabular*}}
\label{tab:Scenarios}
\end{table}

Table~\ref{tab:NuclearDecays} shows the nuclear decay rates corresponding to the different scenarios (table~\ref{tab:Scenarios}) expected in the main spectrometer. The table clearly demonstrates the importance of the NEG strips to reduce the number of tritium $\beta$-decays in the main spectrometer volume as well as the non-negligible number of radon decays even in the case of optimum passive shielding of the pump ports. With an overall decay rate of the order of a few mBq, primarily due to radon emanation from the inner spectrometer walls, a concise calculation of the resulting background rates is mandatory.

\begin{table}
\caption{Expected number of nuclear decays in the main spectrometer for different UHV scenarios. The scenarios are described in detail in the main text.}
\centering
{\footnotesize
\begin{tabular*}{\textwidth}{p{2.5cm}p{2.2cm}p{2.2cm}p{2.2cm}p{2.2cm}}
\hline
& \multicolumn{4}{c}{\textbf{Activity [mBq]}} \\
\hline\hline
& $\text{T}_2$ & ${}^{219}\text{Rn}_{\text{NEG}}$ & ${}^{219}\text{Rn}_{\text{Wall}}$ & ${}^{220}\text{Rn}_{\text{Wall}}$\\
\hline
Scenario 1 & 21.9  & 0 & 27.2 $\pm$ 27.2 & 67.9 $\pm$ 45.0\\
Scenario 2 & 0.01 & 118.7 $\pm$ 29.5 & 27.2 $\pm$ 27.2 & 67.9 $\pm$ 45.0\\
Scenario 3 & 0.02 & 0.12~$\pm$ 0.03 & 5.3~$\pm$ 5.3 & 1.4~$\pm$ 0.1\\
Scenario 4 & 0.05 & 0.05~$\pm$ 0.01 & 5.3~$\pm$ 5.3 & 1.4~$\pm$ 0.1\\
\hline
\end{tabular*}}
\label{tab:NuclearDecays}
\end{table}

\subsection{Expected background rates in different vacuum scenarios}
\label{BackgroundRate}
With the above information on the number of decays as well as the number of secondaries produced in each decay, the total expected background rate can be calculated. The average number of background events $\left\langle N_{B}\right\rangle $ in the energy ROI and in a time interval $t$ longer than the storage time $t>t_\text{s}$ is given by
\begin{eqnarray}
\left\langle N_{B}\right\rangle  = \epsilon_V^\text{MS} \cdot \epsilon_B^\text{MS} \sum_i \left\langle N_{d_i}^\text{MS}\right\rangle \left\langle N_{e_i}^\text{MS}\right\rangle ,
\end{eqnarray}
where $i$ denotes the isotopes tritium and ${}^{219, 220}$Rn, $\left\langle N_{d_i}\right\rangle $ stands for the average number of nuclear decays in a time interval t and $\left\langle N_{e_i}\right\rangle $ represents the average number of electrons produced within one event.

Two further factors in our background estimate have to be taken into account: the sensitive volume of the main spectrometer amounts to only $\epsilon_V^\text{MS} = 0.7$ of the total volume, and only a fraction $\epsilon_B^\text{MS} = 0.4$ of all secondary electrons produced in the main spectrometer will propagate towards the detector (the remaining 60\% fly towards the source side). This asymmetry in the exit direction is due to the asymmetric magnetic field configuration (see figure~\ref{fig:KATRIN}).

Assuming the partial pressures of tritium and radon to be constant over long time periods, the decay rate follows a Poisson distribution. The distribution of the number of secondary electrons is obtained by MC simulations of $10^3$ tritium $\beta$-decays and ${}^{219,220}$Rn $\alpha$-decays each (based on the event generators described in section~\ref{Tools}).

\begin{figure}
\centering
\includegraphics[width=\textwidth]{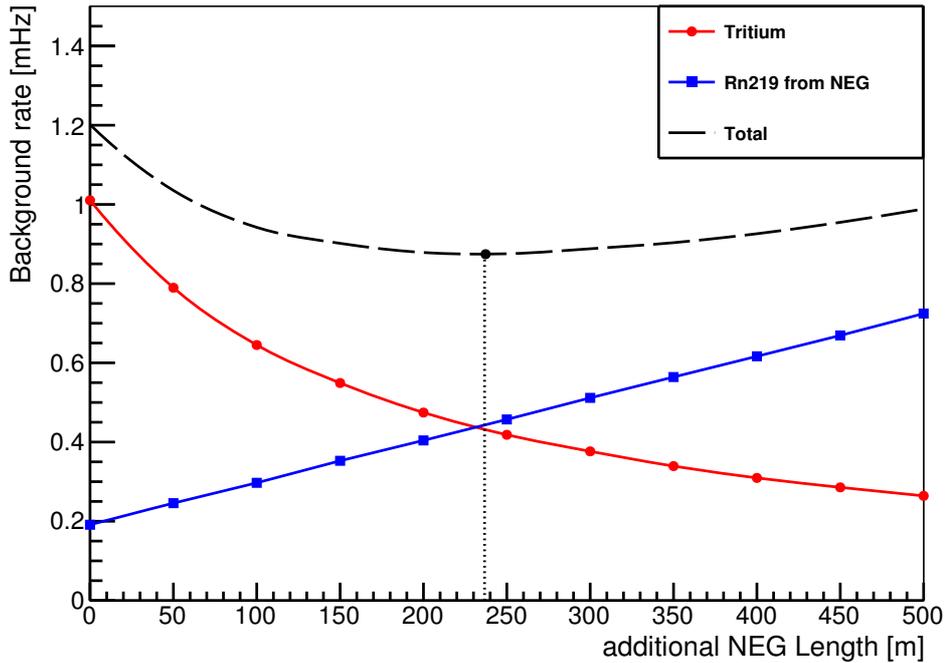}
\caption{\textbf{Estimated contribution of background rate originating from tritium and ${}^{219}$Rn decays as a function of additional getter length in the \prespectrometer{}}. As the plot shows, the radon contribution to the background increases with increasing amounts of getter material whereas the tritium contribution decreases. The optimum for a tritim retention factor as given in~\cite{DesignReport} is found at about 250~m additional getter in the \prespectrometer{} volume. The rather shallow minimum in this case is due to the coincidence of almost identical rates of tritum and radon induced background. In case that the tritium inflow would be larger by an order of magnitude, more NEG strips would be required. The \prespectrometer{} is able to hold up to 1000~m of getter.}
\label{fig:rate_over_neglength}
\end{figure}

On the basis of these considerations, we investigate in detail the influence of specific design modifications on the overall background rate. As an important example we briefly discuss the merits of additional getter strips in the \prespectrometer{}. In figure~\ref{fig:rate_over_neglength}, the trade-off between a reduced background from tritium and an increased radon-induced background due to additional NEG strips is clearly visible.

For the tritium retention factor as stated in \cite{DesignReport}, the tritium-induced background is larger than the ${}^{219}$Rn-induced one by about a factor of five if no additional getter is installed. Were the actual tritium retention factor to differ from the reference value, figure~\ref{fig:rate_over_neglength} would allow adjustments to be made to the \prespectrometer{} getter length.

\begin{figure}
\centering
\includegraphics[width = 1.\textwidth]{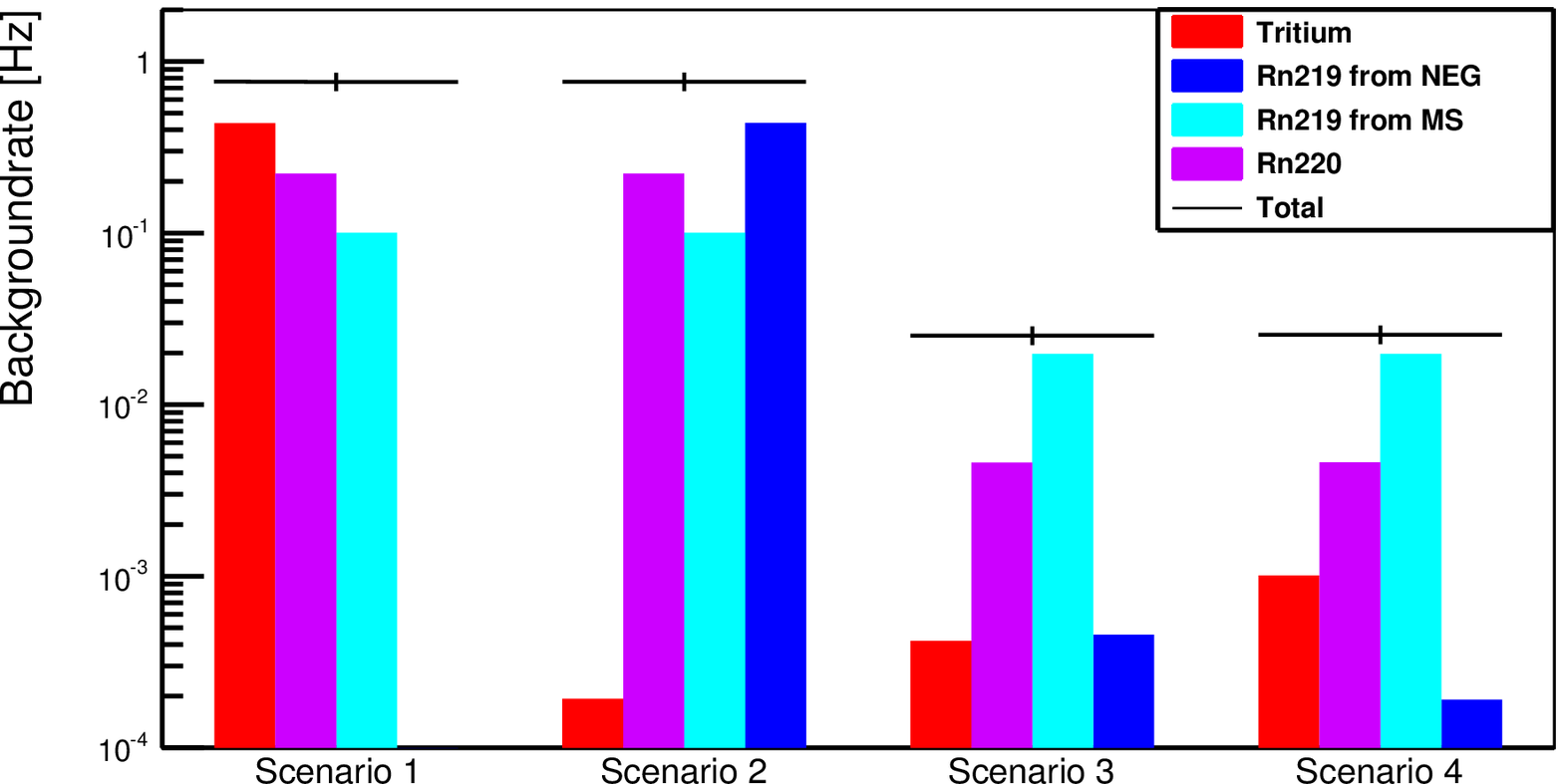}
\caption{\textbf{Expected background rates for four different UHV scenarios described in table~\ref{tab:Scenarios} and section~\ref{DecayRateCalculations}.} In this figure, resulting background rates are broken down into detail. For each scenario, the contribution from each distinct nuclear background source is displayed, illustrating the trade-offs and consequences inherent in implementing the various techniques. Scenarios one and two differ in their use of NEG to combat tritium background at the expense of introducing radon; the similarity of total rates here is purely coincidental. In comparison to scenarios one and two, three and four include a cryo-baffle designed to trap radon emanating from NEG strips.}
\label{fig:barplot}
\end{figure}

These two isotopes, however, are only part of the overall background picture, which is displayed in figure~\ref{fig:barplot} for the four UHV scenarios listed in section~\ref{DecayRateCalculations}. From figure~\ref{fig:barplot} it is evident that scenarios 1 and 2 result in overall background rates of $\sim 1$~Hz, thus exceeding the design criterion of 10~mHz by about 2 orders of magnitude. 

When comparing scenarios 3 and 4 one notes that the total background rate is almost identical. This is because the rate is largely dominated by radon emanation from the inner surface of the main spectrometer, due to the excellent shielding of ${}^{219}$Rn emanation from the pump ports by the $\text{LN}_2$-cooled baffles. However, even when including these passive measures, the expected overall background rate of $\sim 30$~mHz still exceeds the design criterion by a factor of three, pointing to the need for additional active background reduction techniques.  

\section{Impact of the background on the neutrino mass sensitivity of KATRIN}
\label{Sensitivity}
As outlined above, radon emanation from the inner walls of the spectrometer and its structural materials may easily exceed the reference background level by a factor of three, and potentially, in case of larger than expected emanation rates, the background level would be correspondingly larger.

In case of a Poisson-distributed background $\text{N}_{\text{bg}}$, the statistical uncertainty $\sigma_{\text{stat}}$ on the observable $\text{m}_{\overline{\nu}_e}^2$ of KATRIN will scale roughly as $\sim \text{N}^{1/6}_{\text{bg}}$~\cite{DesignReport, Otten}. However, the background arising from stored electrons is not of this type. Instead, the fluctuations of the rate are largely determined by the number of stored primary particles, which is small compared to the number of secondaries reaching the detector. The count rate at the detector therefore shows rather large fluctuations which are not Poisson-distributed. Accordingly, the variance is determined by the variance of the number of stored primaries arising from nuclear decays.

To investigate the impact of the background arising from stored electrons, a detailed model describing the background as a function of time over the full three years measurement time of KATRIN was implemented. The model is based on the full MC simulations described in section~\ref{BackgroundRate} and calculations discussed in section~\ref{DecayRateCalculations}. The statistical sensitivity $\text{m}^{\text{stat}}_{\nu} (90\% \text{C.L.})$ is determined by fitting the theoretical integral $\beta$-spectrum to $10^4$ simulated KATRIN measurements. Each simulation assumes $\text{m}_{\overline{\nu}_e} = 0$~eV and statistically corresponds to three years of data taking. The width of the distribution of the fitted neutrino mass squared $\text{m}^2_{\nu}$  determines the statistical uncertainty $\sigma_{\text{stat}}$ and thereby the neutrino mass sensitivity at 90\% confidence level according to
\begin{equation}
  \text{m}^{\text{stat}}_{\nu} (90\% \text{C.L.}) = \sqrt{1.64 \cdot \sigma_{\text{stat}}}.
\end{equation}
In a typical measurement schedule, the integral tritium $\beta$-spectrum will be measured at 41 different retarding potentials. The overall measurement time at each potential is optimized to achieve the best neutrino mass sensitivity~\cite{DesignReport} for a background level of 10~mHz. During a measurement period of three years, a large number of scans of a few hours' duration through all 41 measurement points will be performed.

\begin{figure}
\begin{center}
\centering
\includegraphics[width = \textwidth]{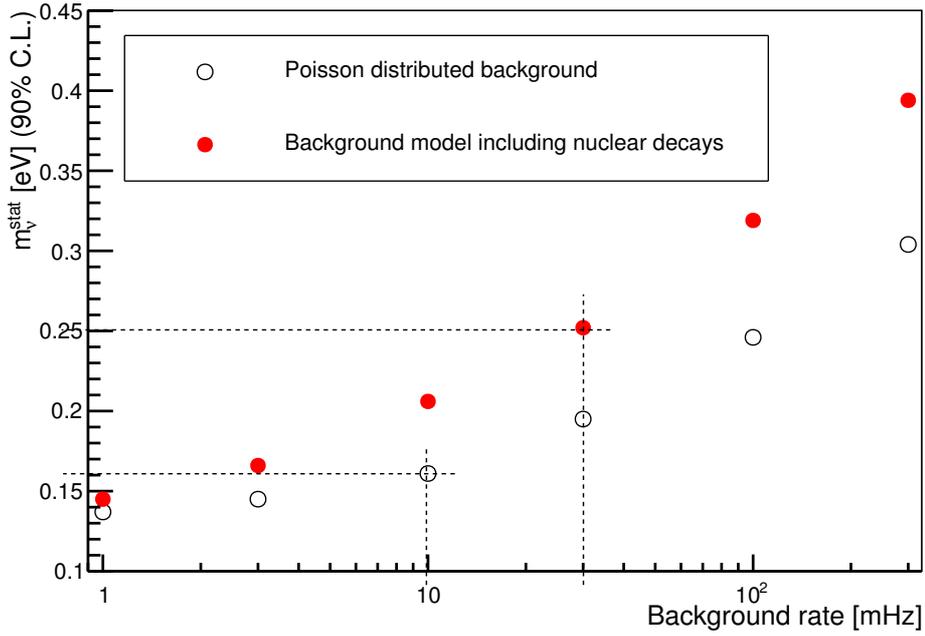}
\caption{\textbf{Statistical neutrino mass sensitivity as a function of background rate for Poisson distributed background and for the background model including nuclear decays}. For these simulations, a fixed scan time of $t_{\text{scan}} = 3$~h and a pressure of $p = 1\cdot10^{-11}$~mbar was used. The dashed lines indicate the statistical sensitivity reached with a Poisson-distributed background of 10~mHz (as stated in~\cite{DesignReport}) and with the estimated background level of 30~mHz (see figure~\ref{fig:barplot}) arising from nuclear decays of this work, if no active reduction methods are implemented.}
\label{fig:sensibackground}
\end{center}
\end{figure}

Figure \ref{fig:sensibackground} shows the statistical neutrino mass sensitivity at 90\% confidence level as a function of the overall background rate (leaving all other contributions at their reference values). Here we compare a Poisson-distributed background (as used in~\cite{DesignReport}) to the background model as calculated in this work including nuclear decays, using a fixed scan time of $t_{\text{scan}} = 3$~h and a pressure level of $\text{p} = 10^{-11}$~mbar.

When comparing both results, it becomes evident that statistical error increases significantly in case of a non-Poissonian background. It is this particular feature of nuclear decays that necessitates the development of active background reduction techniques to realize the full physics potential of KATRIN.

\begin{figure}
\begin{center}
\centering
\includegraphics[width = \textwidth]{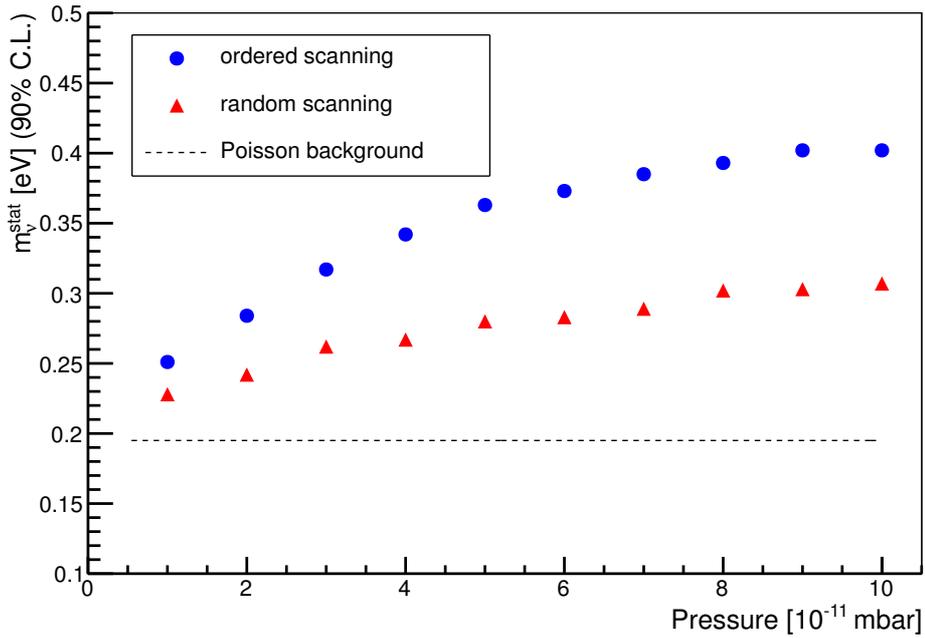}
\caption{\textbf{Statistical neutrino mass sensitivity $\text{m}^{\text{stat}}_{\nu}$ as a function of pressure in the main spectrometer}. For this simulation, a background level of 30~mHz and a scanning time of $t_{\text{scan}} = 3$~h was assumed. For a constant Poisson-distributed background (dashed line) the statistical sensitivity does not depend on the pressure. When the non-Poissonian background arising from nuclear decays is included, the experimental sensitivity $\text{m}^{\text{stat}}_{\nu}$ gets worse for higher pressures, as described in the text. The impact of the non-Poissonian nature of this background can be alleviated by scanning the potentials in random order (triangles), as compared to a fixed order (dots).}
\label{fig:sensipressure}
\end{center}
\end{figure}

\begin{figure}
\begin{center}
\centering
\includegraphics[width = \textwidth]{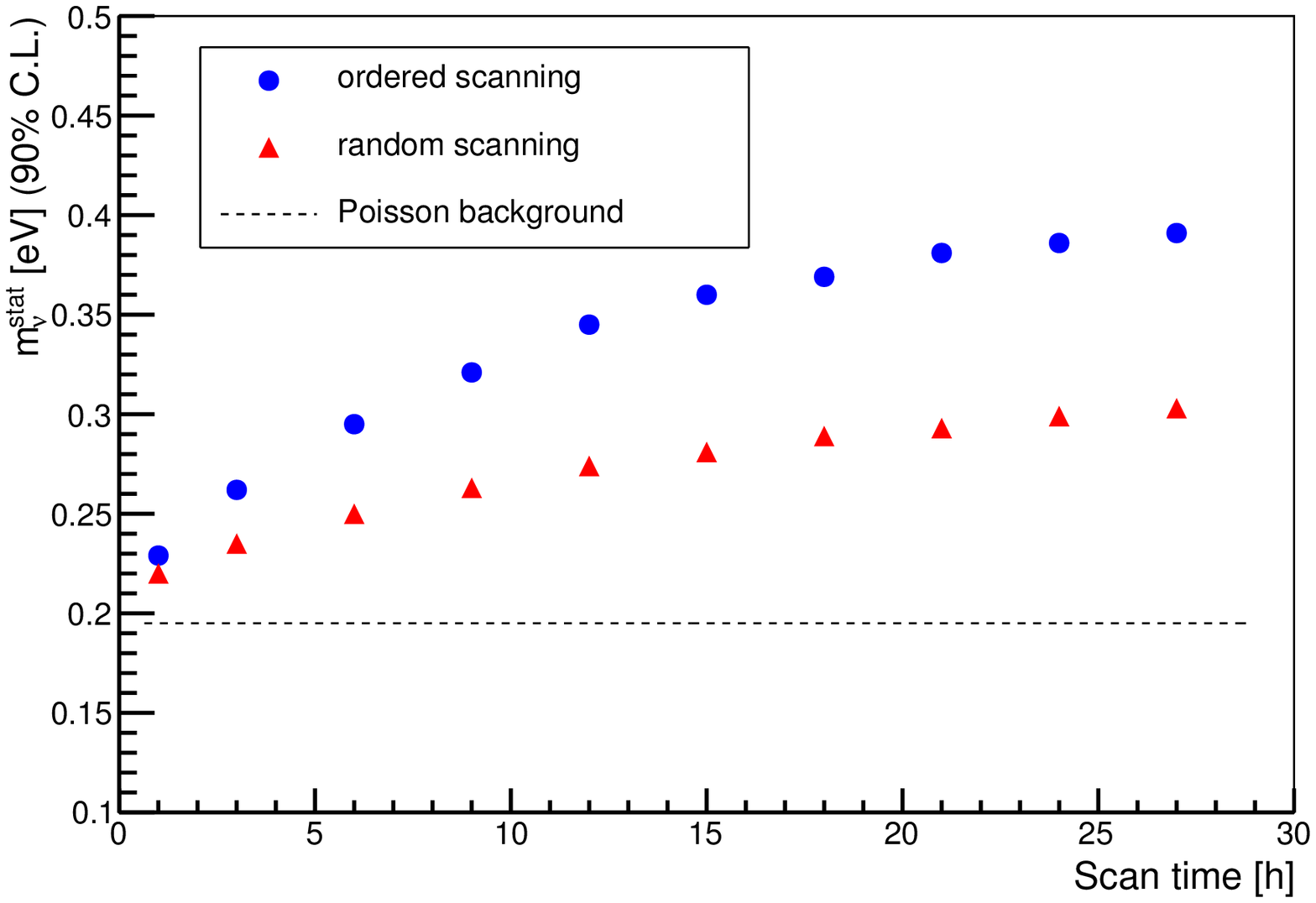}
\caption{\textbf{Statistical neutrino mass sensitivity $\text{m}^{\text{stat}}_{\nu}$ as a function of scanning time}. For this simulation, a background level of 30~mHz and a pressure of $p = 1\cdot10^{-11}$~mbar was used. The presence of a non-Poissonian background results in a dependence of $\text{m}^{\text{stat}}_{\nu}$ on the scanning time in an analogous way as on the pressure, see figure~\ref{fig:sensipressure}. Again, when scanning the potentials in a random way (triangles) the neutrino mass sensitivity is improved, as compared to a scanning method with a fixed order (dots).}
\label{fig:sensiscan}
\end{center}
\end{figure}

Moreover, our simulations revealed that $\text{m}^{\text{stat}}_{\nu} (90\% \text{C.L.})$ strongly depends on the actual time of a scan $t_{\text{scan}}$ and on the total pressure $p$ in the main spectrometer. At scan times $t_{\text{scan}}$ much longer than the storage times $t_{\text{s}}$, the non-Poissonian nature of the background from nuclear decays becomes prominent. Since $t_{\text{s}}$ decreases inverse proportionally with $p$, an analogous effect is observed for higher pressures. Consequently, both larger values of $t_{\text{scan}}$ and larger values of $p$ will result in a decrease of the neutrino mass sensitivity. Furthermore, in an ordered scanning mode, measurements at neighboring filter potentials are correlated due to the long storage times $t_{\text{scan}}$. By scanning the 41 potentials in a random order, this correlation will be alleviated, and consequently the neutrino mass sensitivity can be improved. These interdependencies are visualized in figure~\ref{fig:sensipressure} and~\ref{fig:sensiscan}.   

In summary, our investigations point to the following important facts:
\begin{itemize}
	\item The estimated neutrino mass sensitivity $\text{m}^{\text{stat}}_{\nu}$ has to take into account a detailed background model, the experimental scan mode and the UHV conditions of the spectrometers.
	\item Backgrounds from nuclear decays feature large non-Poissonian rate fluctuations, which result in a decrease of the neutrino mass sensitivity $\text{m}^{\text{stat}}_{\nu}$. 
	\item The neutrino mass sensitivity $\text{m}^{\text{stat}}_{\nu}$ improves with better vacuum, with smaller scanning times, and with randomized scanning (instead of ordered scanning).
\end{itemize}

\section{Conclusion and Outlook}
\label{Conclusion}
Due to their inherent electromagnetic design features, the KATRIN spectrometers act as magnetic bottles for light charged particles. A primary electron in the multi-keV regime produced by a nuclear decay can thus be magnetically trapped over a time period of several hours during which it can produce several hundred secondary electrons.

In this paper, we showed that nuclear decays of tritium migrating from the WGTS to the spectrometer as well as of ${}^{219}$Rn and ${}^{220}$Rn emanating from NEG material and structural components in the volume of the KATRIN main spectrometer can cause a background rate exceeding the design limit of 10~mHz. 

In an optimum scenario, using $\text{LN}_2$-cooled baffles to shield the pump ports, as well as optimized combination of NEG strips and TMPs, a background level of $\sim 30$~mHz is expected.

Of major impact for the neutrino mass sensitivity $\text{m}^{\text{stat}}_{\nu}$ are the large rate fluctuations that this background exhibits. A statistical analysis with a detailed background model revealed a reduction of the statistical neutrino mass sensitivity from $\text{m}^{\text{stat}}_{\nu} = 0.16$~eV at 90\% C.L. (assuming a Poisson distributed background of 10~mHz) to $\text{m}^{\text{stat}}_{\nu} = 0.25$~eV at 90\% C.L. (assuming a realistic background model of 30~mHz).

This result highlights the necessity for further developing active background reduction methods. In a separate publication~\cite{ECR} we describe the successful implementation of the electron cyclotron resonance (ECR) method, which offers great potential in reducing the background described here to a very low level. The upcoming measurements with the KATRIN main spectrometer starting in the second half of 2012 will be of crucial importance to test this and other promising active background reduction methods.

\section*{Acknowledgements}
This work has been supported by the Bundesministerium f{\"u}r Bildung und Forschung (BMBF) with project number 05A08VK2 and the Deutsche Forschungsgemeinschaft (DFG) via Transregio~27 ``Neutrinos and beyond''. We also would like to thank Karlsruhe House of Young Scientists (KHYS) of KIT for their support (S.M., D.F., M.H., W.K., N.W.).

\bibliographystyle{elsarticle-num}
\biboptions{sort&compress}

\end{document}

%% file: abstract.tex
\begin{abstract}
The KATRIN experiment is designed to measure the absolute neutrino mass scale with a sensitivity of 200~meV at 90\% C.L. by high resolution tritium $\beta$-spectroscopy. A low background level of 10~mHz at the $\beta$-decay endpoint is required in order to achieve the design sensitivity.
In this paper we discuss a novel background source arising from magnetically trapped keV electrons in electrostatic retarding spectrometers. The main sources of these electrons are $\alpha$-decays of the radon isotopes ${}^{219,220}$Rn as well as $\beta$-decays of tritium in the volume of the spectrometers. We characterize the expected background signal by extensive MC simulations and investigate the impact on the KATRIN neutrino mass sensitivity. From these results we refine design parameters for the spectrometer vacuum system and propose active background reduction methods to meet the stringent design limits for the overall background rate.
\end{abstract}